\documentclass[conference]{IEEEtran}
\IEEEoverridecommandlockouts

\usepackage{cite}
\usepackage{amsmath,amssymb,amsfonts}
\usepackage{graphicx}
\usepackage{textcomp}
\usepackage{xcolor}
\def\BibTeX{{\rm B\kern-.05em{\sc i\kern-.025em b}\kern-.08em
    T\kern-.1667em\lower.7ex\hbox{E}\kern-.125emX}}
\usepackage[T1]{fontenc}
\usepackage[utf8]{inputenc}
\usepackage{booktabs,multirow,array,tabularx,url}
\usepackage{hyperref}
\newcolumntype{Y}{>{\raggedright\arraybackslash}X}
\begin{document}

\title{SFU-Sim: Bit-Exact Reverse Engineering of NVIDIA MUFU and Validation of PTX Paths}

\author{\IEEEauthorblockN{Haozhi Xie}
\IEEEauthorblockA{
\textit{School of Electronics and Information}\\
\textit{Technology Sun Yat-sen University}\\
Guangdong, China\\
xiehzh23@mail2.sysu.edu.cn}
\and

\IEEEauthorblockN{Tao Su}
\IEEEauthorblockA{
\textit{School of Electronics and Information}\\
\textit{Technology Sun Yat-sen University}\\
Guangdong, China\\
sutao@mail.sysu.edu.cn}}

\maketitle

\begin{abstract}
Reproducing the numerical behavior of a GPU requires predicting the exact output of special function instructions. Vendor documents give only the function semantics and error bounds. These do not fix the output bit pattern. We present SFU-Sim, which builds bit-exact models of NVIDIA multi-function unit (MUFU) instructions by black-box reverse engineering on real GPUs. We first model a single bare MUFU instruction. We then add the surrounding operations found in the disassembly to model full PTX paths. The model covers 12 common MUFU forms and two BF16 forms on the tested Blackwell GPU. Using SIN/COS as a complete case study, we show how we analyze the PTX-to-SASS compilation path and recover the segment structure, the equivalent coefficient tables, and the finite-width arithmetic. On NVIDIA GPUs of three architectures (Ampere, Ada Lovelace, and Blackwell), we exhaustively compare every tested PTX path over its specified input domain. FP16 and FP32 paths cover all $2^{16}$ and $2^{32}$ input encodings, respectively. The model output matches the hardware bit for bit, with zero mismatches. SFU-Sim computes each output directly from its input. It needs neither a lookup table of all inputs nor access to a physical GPU. It can serve as a bit-exact special function component for instruction simulators. It can also serve as an arithmetic reference for hardware design and functional verification that target the same numerical behavior. SFU-Sim and its validation programs are open source at \href{https://github.com/hzX40000000/SFU-Sim}{this https URL}.
\end{abstract}

\begin{IEEEkeywords}
GPU; special functions; MUFU; black-box reverse engineering; bit-exact modeling; exhaustive validation
\end{IEEEkeywords}

\section{Introduction}\label{sec:1}

Instruction-level simulation, numerical reproduction, and low-level arithmetic analysis all need to answer one concrete question: given an instruction and its input bit pattern, what bit pattern does the target GPU return? For IEEE 754 basic operations that require correct rounding, the answer is unique once the operation, operands, target format, and rounding direction are fixed. Reproducing them thus reduces to implementing each operation as the standard defines it. The special function approximation instructions studied in this paper do not have this property. For approximate paths on the special function unit (SFU), NVIDIA usually publishes only function-level error bounds, such as a maximum ULP error or an error range for a given input interval \cite{r1}. An error bound limits how far a result may deviate from a reference value. It does not specify the output bit pattern for each input. Two implementations can meet the same public error bound and still return different bit patterns, because they differ in segment coefficients, intermediate widths, truncation points, or special value encoding. Thus, being closer to the real-valued function is not the same as reproducing the target device. Meeting the public accuracy bound does not mean that one can reproduce a given GPU bit for bit.

The accuracy bound is only the first obstacle. PTX is a virtual instruction interface. The compiler translates it into a machine instruction sequence for the target architecture. A special function PTX instruction may map directly to one MUFU instruction. It may also add input scaling, result correction, or format conversion before or after the MUFU. The result seen by the program depends on the whole sequence, and the bit-level behavior of the MUFU covers only one step. To predict the result, we must describe the numerical behavior of the bare MUFU instruction. We must also identify and model the surrounding operations on the same PTX path. An implementation that only follows the error bounds in the PTX documentation cannot give a definite output bit pattern. Neither can a bare instruction model alone, on paths that need surrounding operations.

Bit-level modeling has prior examples. MMA-Sim builds bit-accurate arithmetic models of matrix multiply-accumulate units across several NVIDIA and AMD architectures through hardware feature probing. It explains cross-platform numerical differences and reveals undocumented precision and rounding behavior \cite{r6}. Its target is matrix multiply-accumulate, and it does not address the piecewise approximation structure of special functions. For special functions, early work involving NVIDIA researchers published evaluation methods and hardware designs based on coefficient tables, quadratic approximation, and dedicated squaring units \cite{r3,r4}. These works do not provide the complete parameters and behavior needed to reproduce the outputs of the devices studied here. For exact reproduction, envytools reads SFU table entries through device registers on early G80/Tesla devices and provides a matching arithmetic implementation \cite{r7}. This table-reading path depends on hardware access specific to old devices. It cannot be applied directly to the modern devices tested in this paper. Cankaya runs all inputs of some MUFU instructions for bit-exact AI inference verification. The hardware outputs are stored in lookup tables of GB scale, so that results can be reproduced without a GPU \cite{r5}. This approach relies on table lookup. It gives no arithmetic model that computes these results directly.

For the modern devices tested here, we need an executable arithmetic model of the bare MUFU. We also need to link this model with the surrounding operations of each PTX path. Together, they enable offline output prediction without a full-input lookup table. To this end, we propose SFU-Sim. It builds bit-level reference models of NVIDIA MUFU instructions through independent black-box reverse engineering. SFU-Sim starts from the minimal PTX paths. It uses SASS to separate the numerical behavior of a single MUFU from the surrounding operations generated by the compiler. We first recover the bare MUFU behavior, bounded by the input and output of one SASS instruction. We then model the surrounding operations and final outputs of the tested PTX paths.

The key difficulty is that the floating-point output encoding does not directly expose the internal fixed-point values or the computation steps. We construct targeted test inputs and analyze the output bit patterns returned by real GPUs. From them, we infer the input reduction, the segment structure, and the coefficient table format. We also convert floating-point observations into allowed integer intervals to recover the coefficients. Using SIN/COS as a complete case study, we show three steps: determining the segment structure and coefficient table format, separating the constant and linear terms, and using quadratic residual analysis to recover the quadratic coefficients and the truncated partial-product squaring rule.

We implement the SFU-Sim model in C++ and validate it against real GPUs with CUDA programs. The model covers eight FP32 forms (EX2, LG2, SIN, COS, TANH, RCP, RSQ, and SQRT), two FP16 forms (EX2 and TANH), and two FP64 high-word forms (RCP64H and RSQ64H). It also supports EX2.BF16 and TANH.BF16 on the tested Blackwell device. In addition, SFU-Sim supports the tested non-\texttt{.ftz} FP32 approximate paths, packed FP16x2 paths, and FP32 RCP and SQRT paths with the RN, RZ, RM, and RP rounding modes. On target devices of three architectures, we exhaustively compare each tested path over its specified input domain. The model output matches the hardware bit for bit, with zero mismatches. Section~\ref{sec:4} gives the validation scope for each device and path.

SFU-Sim changes how the special function outputs of a GPU are reproduced. Instead of relying on a physical device or a full-input lookup table, it computes outputs on demand from a finite set of coefficients and arithmetic rules. It can serve as a bit-level numerical component for instruction simulators. It can also serve as a reference specification for SFU hardware design and functional verification that target the same numerical behavior. Our main contributions are as follows.
\begin{enumerate}
\setlength{\itemsep}{1pt}
\item \textbf{Bit-level arithmetic models of the target MUFU instructions.} Through independent black-box reverse engineering, we recover arithmetic models for the 12 common forms and the two BF16 forms on the tested Blackwell GPU. We clearly define the boundary between bare MUFU behavior and the surrounding PTX operations.
\item \textbf{A method to recover finite arithmetic from hardware observations.} Using the SIN/COS instructions, we show the full reverse engineering process: separating the surrounding instructions, determining the segments and the coefficient table format, applying integer interval constraints, separating the parameters, and analyzing the quadratic residuals.
\item \textbf{Multi-device validation and offline output prediction.} We run exhaustive comparisons over the specified input domains of the tested PTX paths on three target devices. The model matches the hardware bit for bit. We also provide function interfaces to predict the outputs of the tested PTX paths.
\end{enumerate}

\section{Background and Problem Definition}\label{sec:2}

\subsection{PTX, SASS, and the Bare MUFU Boundary}\label{sec:2.1}

PTX is the virtual instruction set of NVIDIA GPUs. After compilation or just-in-time compilation, a PTX program becomes machine instructions for the target architecture. These machine instructions can be shown as SASS by disassembly and inspected with NVIDIA binary tools \cite{r1,r2}. PTX and SASS instructions do not always map one to one. For a given target architecture and compile setting, a special function PTX instruction may map to a single MUFU instruction. It may also map to a SASS sequence that includes other surrounding instructions. For example, \texttt{sin.\allowbreak{}approx.\allowbreak{}ftz.\allowbreak{}f32} compiles to an \texttt{FMUL.\allowbreak{}RZ} prescale followed by \texttt{MUFU.\allowbreak{}SIN} (Table~\ref{tab:1}). We call the complete implementation of one PTX special function instruction on a given device and compile setting a \textit{PTX path}.

We identify the target operation by the \texttt{MUFU.\ensuremath{*}} instruction in SASS. The register input-output relation of one such instruction is the observation boundary of the bare model. The model maps the bit pattern in the source register to the bit pattern in the destination register. It includes how the instruction itself encodes zeros, subnormals, infinities, and NaNs. Other SASS instructions before or after the MUFU, such as input scaling, result correction, and format conversion, lie outside this boundary.

To reproduce the full observable result of a PTX path on the GPU, we build software models that are functionally equivalent to the surrounding operations. These models follow the actual SASS sequence from disassembly. They cover input preprocessing, result postprocessing, and format conversion. We then combine them with the bare MUFU model. For paths built on bare MUFU behavior, SFU-Sim has two parts. The bare MUFU model describes the bit-level behavior of a single MUFU instruction. The PTX path model reproduces the surrounding operations before and after it. The combined result is compared bit for bit with the end-to-end output of the GPU. For the RCP/SQRT paths with a specified final rounding, SFU-Sim instead evaluates the final output semantics defined by PTX (Section~\ref{sec:4.3}).

\subsection{Study Scope}\label{sec:2.2}

We first choose PTX forms that have as few surrounding operations as possible and whose surrounding operations can be identified and removed through disassembly. This reduces the ambiguity in whether a final output comes from the bare MUFU or from the surrounding operations. Table~\ref{tab:1} lists the minimal entries used, their target MUFU instructions, and their surrounding operations. ``None'' means that the final result directly constrains the bare model.
\begin{table*}[!t]
\caption{Minimal PTX entries and their target MUFU instructions}\label{tab:1}
\centering\footnotesize
\setlength{\tabcolsep}{3pt}\renewcommand{\arraystretch}{1.1}
\begin{tabularx}{\textwidth}{@{}l l Y@{}}
\toprule
\textbf{PTX entry} & \textbf{Target MUFU} & \textbf{Key surrounding instructions and their roles} \\
\midrule
\texttt{ex2.\allowbreak{}approx.\allowbreak{}ftz.\allowbreak{}f32} & \texttt{MUFU.\allowbreak{}EX2} & None \\
\texttt{lg2.\allowbreak{}approx.\allowbreak{}ftz.\allowbreak{}f32} & \texttt{MUFU.\allowbreak{}LG2} & None \\
\texttt{rcp.\allowbreak{}approx.\allowbreak{}ftz.\allowbreak{}f32} & \texttt{MUFU.\allowbreak{}RCP} & None \\
\texttt{rsqrt.\allowbreak{}approx.\allowbreak{}ftz.\allowbreak{}f32} & \texttt{MUFU.\allowbreak{}RSQ} & None \\
\texttt{sqrt.\allowbreak{}approx.\allowbreak{}ftz.\allowbreak{}f32} & \texttt{MUFU.\allowbreak{}SQRT} & None \\
\texttt{tanh.\allowbreak{}approx.\allowbreak{}f32} & \texttt{MUFU.\allowbreak{}TANH} & None \\
\texttt{sin.\allowbreak{}approx.\allowbreak{}ftz.\allowbreak{}f32} & \texttt{MUFU.\allowbreak{}SIN} & \texttt{FMUL.\allowbreak{}RZ} prescale \\
\texttt{cos.\allowbreak{}approx.\allowbreak{}ftz.\allowbreak{}f32} & \texttt{MUFU.\allowbreak{}COS} & \texttt{FMUL.\allowbreak{}RZ} prescale \\
\texttt{ex2.\allowbreak{}approx.\allowbreak{}f16} & \texttt{MUFU.\allowbreak{}EX2.\allowbreak{}F16} & None \\
\texttt{tanh.\allowbreak{}approx.\allowbreak{}f16} & \texttt{MUFU.\allowbreak{}TANH.\allowbreak{}F16} & None \\
\texttt{rcp.\allowbreak{}approx.\allowbreak{}ftz.\allowbreak{}f64} & \texttt{MUFU.\allowbreak{}RCP64H} & \texttt{LDG.\allowbreak{}E} loads the high 32 bits of the input into the MUFU; \texttt{HFMA2.\allowbreak{}MMA} sets the low 32 bits of the output to zero \\
\texttt{rsqrt.\allowbreak{}approx.\allowbreak{}ftz.\allowbreak{}f64} & \texttt{MUFU.\allowbreak{}RSQ64H} & \texttt{LDG.\allowbreak{}E} loads the high 32 bits of the input into the MUFU; \texttt{HFMA2.\allowbreak{}MMA} sets the low 32 bits of the output to zero \\
\texttt{ex2.\allowbreak{}approx.\allowbreak{}ftz.\allowbreak{}bf16}\textsuperscript{\textdagger} & \texttt{MUFU.\allowbreak{}EX2.\allowbreak{}BF16} & \texttt{LOP3.\allowbreak{}LUT} masks off the high 16 bits of the output \\
\texttt{tanh.\allowbreak{}approx.\allowbreak{}bf16}\textsuperscript{\textdagger} & \texttt{MUFU.\allowbreak{}TANH.\allowbreak{}BF16} & \texttt{LOP3.\allowbreak{}LUT} masks off the high 16 bits of the output \\
\bottomrule
\end{tabularx}
\par\smallskip{\footnotesize\raggedright\textbf{Note:} \textsuperscript{\textdagger} Available only on the tested Blackwell (\texttt{sm\_\allowbreak{}120}). The other 12 entries are common to all three devices. The last column lists the surrounding SASS operations that appear on each path.\par}
\end{table*}

The first 12 entries form the common minimal set. It contains eight FP32 forms, two FP16 forms, and two FP64 high-word forms. Each is validated on all three target devices.

Beyond the minimal entries, SFU-Sim also covers non-\texttt{.ftz} FP32 approximate paths, packed FP16x2 paths, and FP32 RCP/SQRT paths with the RN, RZ, RM, and RP rounding modes. Section~\ref{sec:4.3} describes how these paths are predicted and validated.

\subsection{Consistency Criterion}\label{sec:2.3}

For every input in the input domain, we compare the full output bit pattern of the software with that of the real GPU. Any differing bit counts as a mismatch. Validation requires zero mismatches. Thus, differences in signed zeros, NaN encodings, infinities, or any other output bit all count as mismatches. They are not excused because the values are close.

The input domain follows the input format of each path. Scalar FP32 paths cover all $2^{32}$ encodings. Scalar FP16/BF16 paths cover all $2^{16}$ encodings. Packed FP16x2 paths cover all combinations of the two half-words. On the FP64 high-word paths, \texttt{MUFU.\allowbreak{}RCP64H}/\texttt{MUFU.\allowbreak{}RSQ64H} read only the high 32 bits of the input, and a surrounding instruction sets the low 32 bits of the output to zero (Table~\ref{tab:1}). The low input word does not take part in the computation of these paths. The experiments therefore fix the low word to zero, cover all $2^{32}$ high words, and compare the full 64-bit output. This domain covers all inputs that the path can distinguish, but it is not the full set of $2^{64}$ FP64 encodings.

\section{Black-Box Reverse Engineering and Finite-Arithmetic Model Construction}\label{sec:3}

\subsection{Overall Approach}\label{sec:3.1}

Published design work describes the basic organization of such units. The reduced input is split into high bits and low bits. The high bits serve as a segment index that selects a set of coefficients. The low bits form the in-segment coordinate $X_2$. Within each segment, the unit performs quadratic interpolation as
\begin{equation}\label{eq:1}
C_0+C_1X_2+C_2X_2^2.
\end{equation}

A dedicated squaring unit generates the quadratic term. The coefficients are stored with finite word lengths. When the coefficient range allows, some leading bits are not stored and are added back at the table output \cite{r4}. That work gives configurations with 32, 64, and 128 segments, which we use as the first segment counts to test. These configurations take the true mathematical value as the reference, and the designers choose them by trading area against accuracy. Our criterion is different. We must reproduce the actual output bit pattern of the target GPU, not approximate the true value.

\begin{figure*}[!t]
\centering
\includegraphics[width=0.74\textwidth]{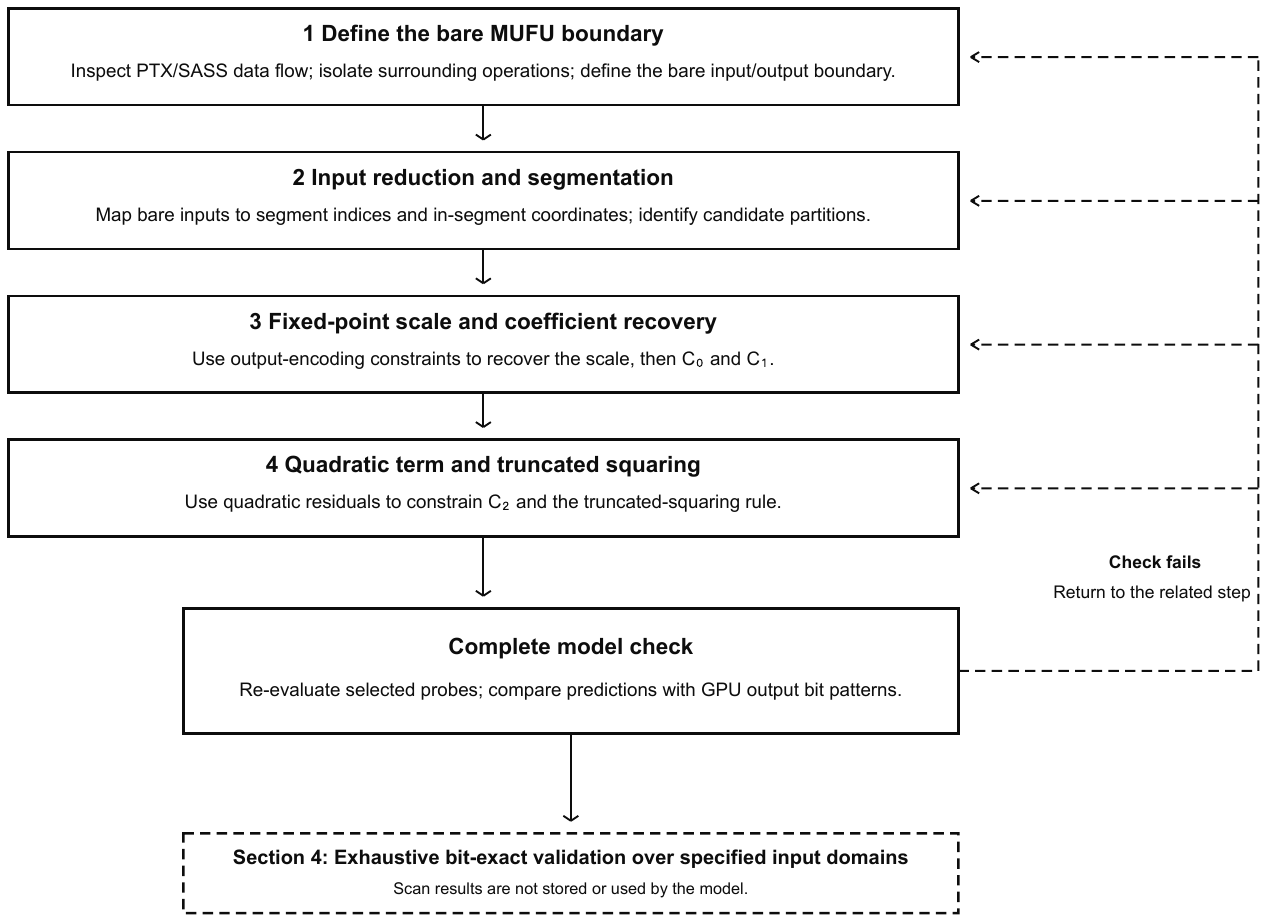}
\caption{Reverse engineering flow and check against hardware.}\label{fig:1}
\end{figure*}

Published work offers candidate organizations, but it cannot fix the exact design of the tested GPUs. For example, where is the boundary between the bare MUFU and the surrounding PTX operations? How does the bare input form the segment index and the in-segment coordinate, and what segment structure is used? How is the finite arithmetic in each segment organized, including the fixed-point scale, the coefficient format, and the way the square term is produced? We recover these in the order shown in Fig.~\ref{fig:1}. First, we identify and separate the surrounding PTX operations to fix the input-output boundary of the bare MUFU. Next, we analyze how the input reduction maps the bare input to a segment index and an in-segment coordinate, and from this we determine the segment structure. We then recover the fixed-point scale and the constant and linear coefficients. Finally, we jointly constrain the quadratic coefficients and the squaring rule, and check the complete model against the hardware. Each step has a falsifiable test on the observations. If a test fails, we return to the related hypothesis. One principle runs through this section. We first check whether a candidate agrees with the observed hardware outputs. Only when several candidates are equivalent on the existing observations do we choose among them by structural compactness and design logic. A smaller fitting error never replaces bit-level agreement. Section~\ref{sec:3.2} shows the full reverse engineering process with SIN/COS as an example. Section~\ref{sec:3.3} summarizes the recovered finite-arithmetic MUFU models. It covers the overall model structure, the shared arithmetic, and the input reduction and special handling of each function.

\subsection{Reverse Engineering and Model Construction for SIN/COS}\label{sec:3.2}

SIN/COS involve surrounding PTX operations, phase quantization, quadrant reuse, and piecewise quadratic evaluation all at once. We use them to show the recovery process from the bare MUFU boundary to a finite-arithmetic model. Section~\ref{sec:3.3} describes how the other forms differ.

\subsubsection{Removing Surrounding Operations: Defining the Boundary}\label{sec:3.2.1}

Reverse engineering SIN/COS first requires the bare MUFU boundary, because the PTX input is not the bare core input. On the tested minimal path, the disassembly is
\begin{flushleft}\ttfamily\footnotesize
FMUL.RZ\ \ t, x, 0x3e22f983\ \ ; 0x3e22f983 = FP32($1/2\pi$)\\
MUFU.SIN\ \ y, t
\end{flushleft}

The multiplication is an FP32 multiply that rounds toward zero. The multiplier is the FP32 value of $1/2\pi$. It converts radians into turns $t$. Besides this multiply, the tested path has no other numerical or format operation. The bare MUFU analysis below takes $t$ as its input.

\subsubsection{Phase Representation and Quadrant Reuse}\label{sec:3.2.2}

For a normal finite bare MUFU input $t$, the model extracts the phase within one period from $|t|$. The input sign is kept separately and used to restore the output sign as each function requires. Sine and cosine repeat every whole turn, and the measured outputs follow this relation. The phase analysis thus only needs the fractional part $f\in[0,1)$ of $|t|$. The number of whole turns plays no role in later steps.

Evidence from two directions bounds the number of effective phase bits. Some input pairs share their first 24 bits but give different outputs, so at least 25 bits are used. Input pairs that share their first 25 bits always give the same output, so lower bits do not take part, and at most 25 bits are used. Together, these show that the in-period phase has exactly 25 bits.

Let this 25-bit code be $f_{25}=\lfloor2^{25}f\rfloor\bmod2^{25}$. Its top 2 bits are the quadrant $q$, and its low 23 bits are the in-quadrant phase $p$. By trigonometric symmetry, every output equals the magnitude kernel $K$ evaluated at $p$ or at the mirrored phase $p'$, multiplied by a sign. The quadrant and the function together decide which phase and which sign to use (Table~\ref{tab:2}). There are two candidates for the mirror. One is the endpoint reflection $2^{23}-p$ of the 23-bit phase code. The other is the bitwise inversion $p'=\mathtt{0x7FFFFF}-p$. They differ by one least significant bit. Only the latter matches all measured outputs bit for bit, so the mirror is implemented by bitwise inversion.
\begin{table}[!t]
\caption{Phase selection and sign in the four quadrants}\label{tab:2}
\centering\small
\setlength{\tabcolsep}{3pt}\renewcommand{\arraystretch}{1.1}
\begin{tabular}{@{}l c c@{}}
\toprule
\textbf{Quadrant} & \textbf{sin} & \textbf{cos} \\
\midrule
First & $+K(p)$ & $+K(p')$ \\
Second & $+K(p')$ & $-K(p)$ \\
Third & $-K(p)$ & $-K(p')$ \\
Fourth & $-K(p')$ & $+K(p)$ \\
\bottomrule
\end{tabular}
\par\smallskip{\footnotesize\raggedright\textbf{Note:} $p'=\mathtt{0x7FFFFF}-p$, that is, the bitwise inverse of the in-quadrant phase.\par}
\end{table}

This mapping expresses four quadrants and two functions as one magnitude kernel evaluated at different phases with different signs. After this reduction, what looks like ``4 quadrants $\times$ 2 functions = 8 behaviors'' becomes a single-kernel problem with a 23-bit phase input. The following sections deal only with this kernel. Quadrant recovery and sign handling are treated as operations outside the kernel.

\subsubsection{Segment Structure}\label{sec:3.2.3}

From Table~\ref{tab:2}, the 23-bit phase code fed to the magnitude kernel $K$ is $p$ or its mirror $p'$. We call it \texttt{phase23}. Its high bits select the segment coefficients, and its low bits form the in-segment coordinate. Based on the candidate structures in Section~\ref{sec:3.1}, we compare three partitions: 32, 64, and 128 segments. The overall fitting error under different segment counts cannot serve as the criterion directly. With more segments, each segment covers a shorter phase range, and a quadratic curve usually fits the observations more easily. A lower error therefore does not mean that the boundaries are more accurate. To locate the boundaries, we instead compare the fitting errors of equal-length candidate intervals that are paired from different starting points.

\begin{figure*}[!t]
\centering
\includegraphics[width=0.82\textwidth]{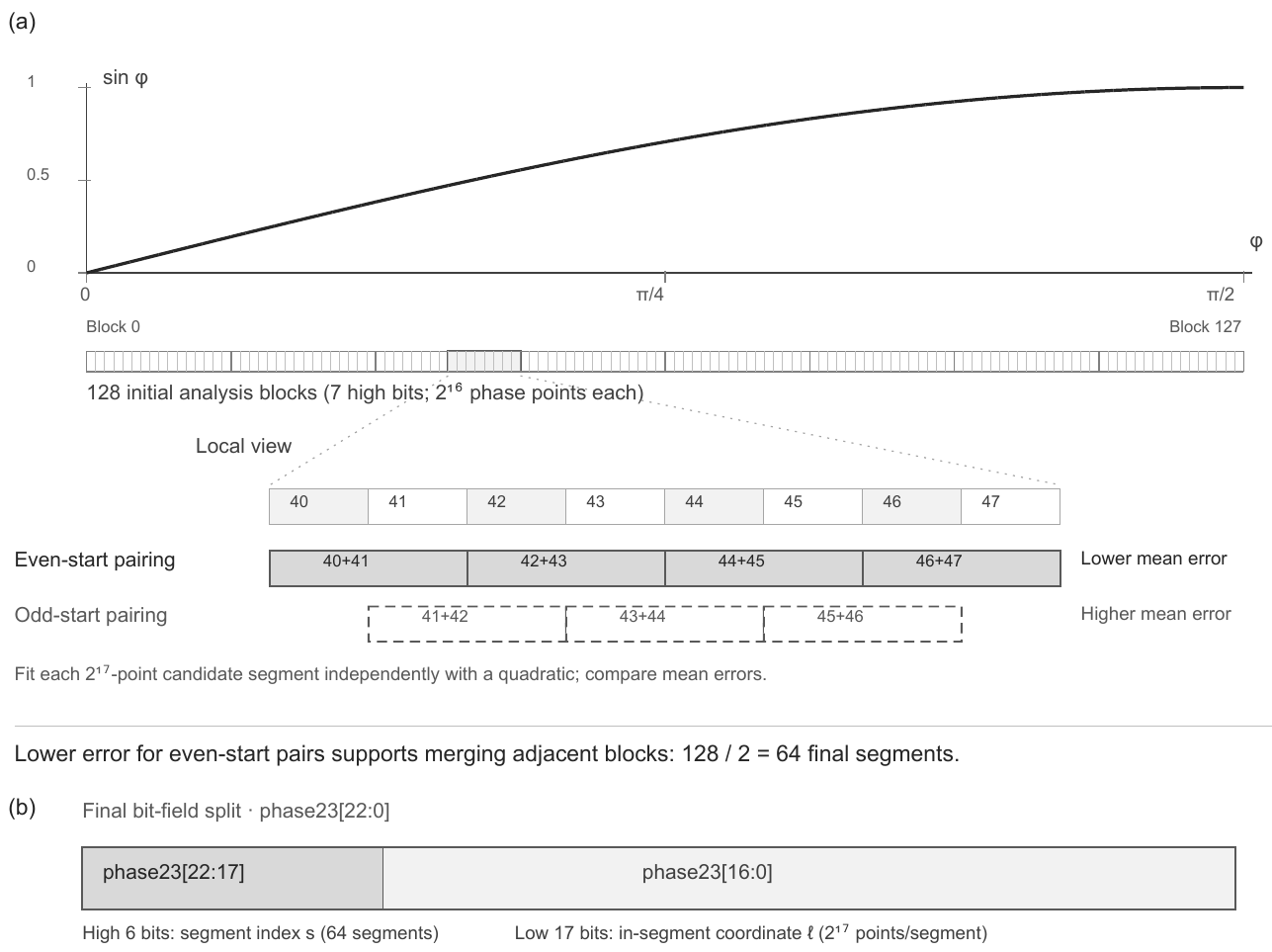}
\caption{Segment detection and bit-field split. (a) Fitting errors of even-start and odd-start pairings; (b) the 6+17-bit split for 64 segments.}\label{fig:2}
\end{figure*}

The three partitions give segment lengths of $2^{18}$, $2^{17}$, and $2^{16}$ phase points. We take $2^{16}$ points as one base block and join two adjacent blocks into a candidate interval of $2^{17}$ points. Pairing starts either at even blocks or at odd blocks (Fig.~\ref{fig:2}a). We fit a quadratic to each candidate interval and compare the fitting errors of the two pairings.

Where the two starting points fall relative to the true segment boundaries depends on the real segment length. A candidate interval that crosses a boundary may contain two different in-segment approximations, one from each side. A single quadratic curve therefore fits it worse than an interval that lies inside one segment. Table~\ref{tab:3} lists what each of the three hypotheses predicts.
\begin{table*}[!t]
\caption{Boundary positions and pairwise error patterns under the three segment-count hypotheses}\label{tab:3}
\centering\footnotesize
\setlength{\tabcolsep}{3pt}\renewcommand{\arraystretch}{1.1}
\begin{tabularx}{\textwidth}{@{}l Y Y Y@{}}
\toprule
\textbf{Hypothesized segments} & \textbf{Even-start pairs} & \textbf{Odd-start pairs} & \textbf{Expected pairwise error pattern} \\
\midrule
32 & All inside true segments & Every other pair crosses a true boundary & Larger odd-start errors appear in every other pair, in an alternating pattern \\
64 & All inside true segments & Every pair crosses a true boundary & Even-start errors stay below odd-start errors for every pair \\
128 & Every pair crosses a true boundary & Every pair crosses a true boundary & Neither start shows a consistently lower error \\
\bottomrule
\end{tabularx}
\end{table*}

In the measurements, the even-start errors stay below the odd-start errors, and the odd-start errors do not rise in every other pair. This pattern matches the 64-segment prediction in Table~\ref{tab:3} and does not match the 32- or 128-segment predictions. We therefore adopt 64 segments. We split \texttt{phase23} into a 6-bit high segment index $s$ ($s=0,\ldots,63$) and a 17-bit low in-segment coordinate $\ell$. Each segment thus covers $2^{17}$ phase points (Fig.~\ref{fig:2}b).

\subsubsection{Fixed-Point Scale and Linear Coefficients}\label{sec:3.2.4}

With the split of the 23-bit phase fixed, the next step is to find the scale of the internal fixed-point representation. We base it on the constraints that the measured outputs place on internal values. Let $i$ index the observation points and let $o_i$ be the FP32 output at point $i$. Two common rules convert an internal fixed-point value to FP32: truncation and round-to-nearest. The intervals they give differ by half a least significant bit. The derivation below assumes truncation. Every internal value between $o_i$ and its next representable value $\mathrm{next}(o_i)$ gives the same bit pattern. So one measured output only limits the internal value to the interval $[o_i,\mathrm{next}(o_i))$. It cannot fix the exact value. Suppose the internal fixed-point value uses a Q format with $Q$ fractional bits, so that the internal value is $y/2^Q$ with $y\in\mathbb{Z}$. The output can then be represented if and only if an integer $y$ satisfies
\begin{equation}\label{eq:2}
\begin{gathered}
\exists\, y\in\mathbb{Z}:\ L_i\le y\le H_i,\\
L_i=\left\lceil o_i\,2^{Q}\right\rceil,\ \ H_i=\left\lceil \mathrm{next}(o_i)\,2^{Q}\right\rceil-1 .
\end{gathered}
\end{equation}

$L_i\le H_i$ means that a feasible integer exists. $L_i>H_i$ means that this scale cannot produce this output.

The scale must work for all observation points at the same time. Each extra bit of $Q$ makes the internal grid twice as fine, and outputs that were representable remain representable. Each observation point therefore has a single threshold $Q_{\min}(i)$, and any $Q\ge\max_i Q_{\min}(i)$ satisfies all points. Over all $2^{23}$ phase points, this maximum is 37. At $Q=36$, at least one point has an empty interval. At $Q=37$, no interval is empty. A larger $Q$ also works, but it only shifts the same solution to the left and widens the datapath and the coefficient storage. The model therefore uses the smallest feasible scale, Q37.

Once Q37 is fixed, the width of each allowed interval equals the gap between adjacent FP32 values times $2^{37}$. This width changes with the output magnitude. For small outputs, the interval can shrink to a single integer. Near the start of a segment, the quadratic term also contributes little. We can therefore test whether a linear model alone satisfies the output constraints at these points. The measurements show that, at the start of each segment, one straight line passes through the allowed intervals of all tested points. For segment $s$, let the constant and linear coefficients be $c_s$ and $b_s$, and set the integer in Eq.~\eqref{eq:2} to $y=c_s+b_s\ell_i$. Each point requires $y$ to fall in its allowed interval, that is,
\begin{equation}\label{eq:3}
\max_i\!\left(L_i-b_s\ell_i\right)\le c_s\le \min_i\!\left(H_i-b_s\ell_i\right).
\end{equation}

A nonempty intersection means that an integer pair $(c_s,b_s)$ exists. The coefficient format and the intersection of the constraints from multiple starting points shrink the feasible set to a unique solution. Solving the 64 segments separately gives one pair $(c_s,b_s)$ per segment. These pairs serve as the linear baseline for the later fitting.

The range in which the linear model holds is found by measurement. We check point by point whether the line still falls in each point's allowed interval. In every segment, this holds for the first 1024 points. Beyond them, the deviation leaves the allowed interval, and a quadratic term is needed. With this baseline, the remaining deviation determines the quadratic term and how the square is implemented.

\subsubsection{Quadratic Term and Square Generation}\label{sec:3.2.5}

For segment $s$, we keep the notation of Section~\ref{sec:3.2.4}. Here $\ell_i$ is the in-segment coordinate, and $c_s$ and $b_s$ are the integer constant and linear coefficients already found for this segment. Subtracting the linear term $c_s+b_s\ell_i$ from the allowed integer interval $[L_i,H_i]$ of each point gives the residual interval $R_i$ that the quadratic term must satisfy:
\begin{equation}\label{eq:4}
R_i=\bigl[L_i-(c_s+b_s\ell_i),\ H_i-(c_s+b_s\ell_i)\bigr].
\end{equation}

We first test square forms without correction. The candidate parameters are the number of bits to which the input is quantized before squaring, the number of bits to which the square is quantized, and the rounding rule. With no input quantization, the candidate is the full square. For each candidate, we check whether a single integer quadratic coefficient $a_s$ for the segment puts the quadratic term of every point inside its residual interval $R_i$. Within the parameter ranges that we explored, no uncorrected candidate satisfies all constraints of the tested segments. A single quantized main square term is therefore not enough to explain the observations.

We then allow the main square term to carry a correction function shared by all segments. Among the combinations that we explored, take the square input $v=\ell_i\gg2$ and the full-square main term $v^2\gg15$. Using the residual intervals $R_i$ over the full input range of each segment, we can jointly solve for the quadratic coefficient $a_s$ of segment $s$ and the shared correction $\delta(v)$. The square term becomes $(v^2\gg15)-\delta(v)$, where $\delta(v)\in\{0,1,\ldots,7\}$ depends only on the low 14 bits of $v$. This form reproduces the observed first-quadrant outputs bit for bit. Implemented this way, however, it needs an extra correction table indexed by the low 14 bits of $v$.

The recovered correction $\delta(v)$ is always non-negative and small, and it changes only with the low bits of the square input. This suggests that the target square term equals the full square minus a small amount set by the low bits. A square can be written as a sum of non-negative partial products, grouped by bit weight. The low-weight columns involve only the low input bits, and their weights limit their contribution. These features give a testable structural explanation: the target squaring may omit some low-weight partial products.

\begin{figure*}[!t]
\centering
\includegraphics[width=0.75\textwidth]{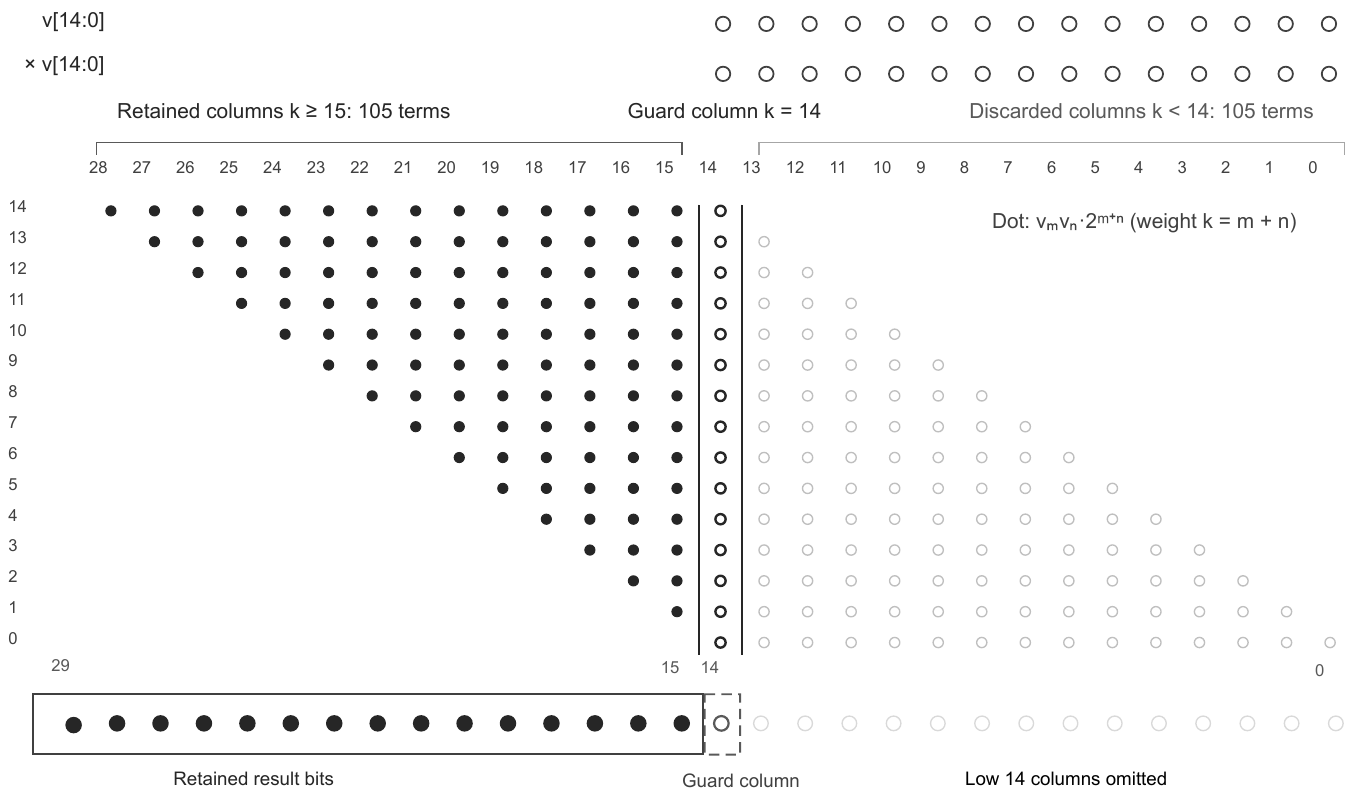}
\caption{Partial-product structure of the truncated square. Columns with $m+n\ge14$ are kept. Column 14 does not enter the result directly, but its carry can affect column 15.}\label{fig:3}
\end{figure*}

Bit $m$ and bit $n$ of the square input form a partial product of weight $2^{m+n}$. To find which products are omitted, we keep the partial products starting from different weight columns. For each choice, we check whether the difference between the truncated square and the full square, after a 15-bit right shift, equals the $\delta(v)$ derived from the observations. When we sum the partial products with $m+n\ge14$, this difference equals $\delta(v)$ at every one of the $2^{15}$ square inputs. Column 14 does not enter the shifted result directly, but its carry can still affect column 15. The equivalent rule therefore keeps it. We call the square given by this rule the truncated square $S_{15}(v)$. Figure~\ref{fig:3} shows the corresponding partial-product region.

The correction table can therefore be replaced by this truncated partial-product rule. Combining the rule with the recovered segment coefficients gives a model that reproduces the measured first-quadrant SIN/COS outputs bit for bit.

\subsubsection{Complete Equivalent Model}\label{sec:3.2.6}

\begin{figure*}[!t]
\centering
\includegraphics[width=0.75\textwidth]{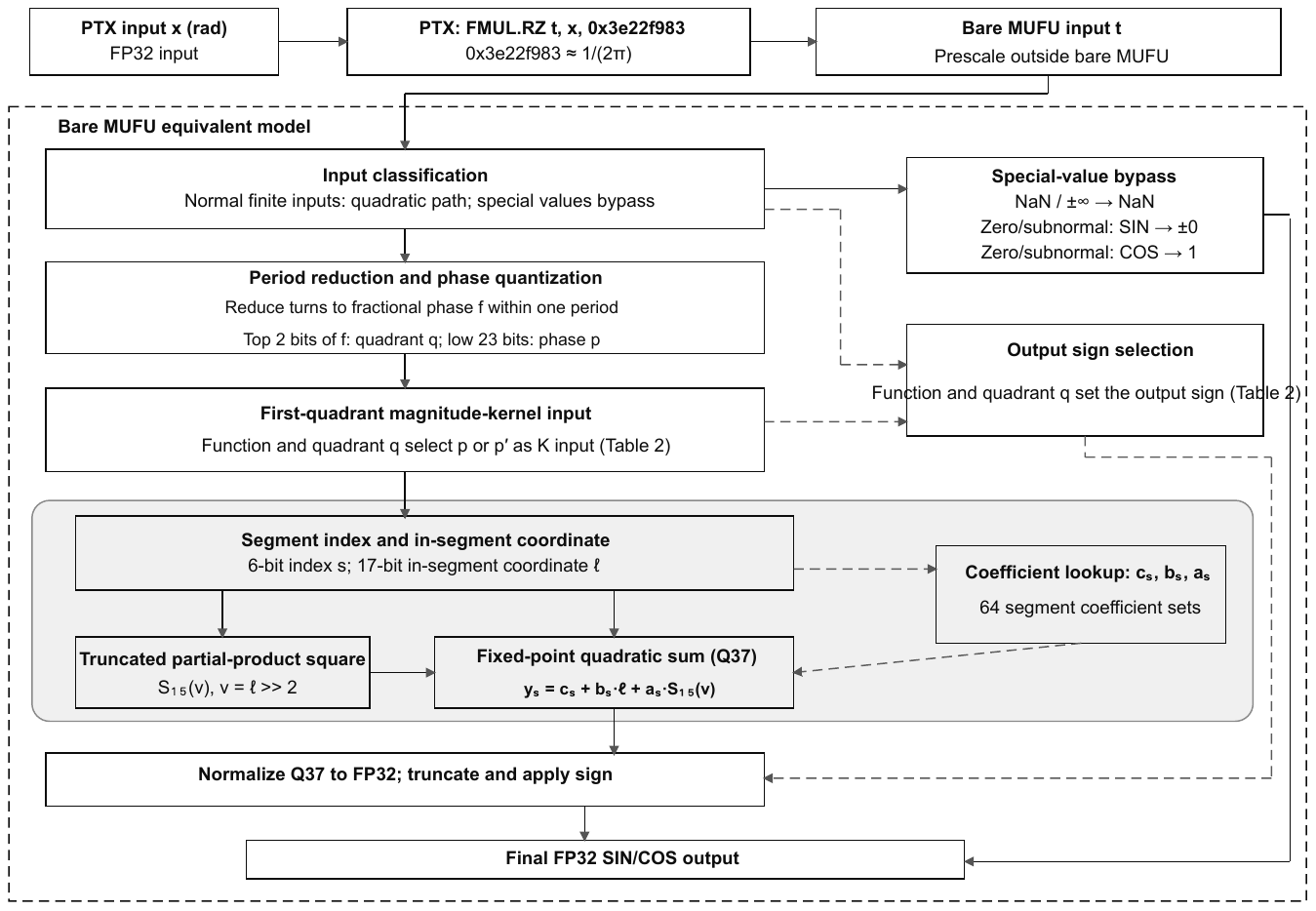}
\caption{Data flow of the equivalent SIN/COS model.}\label{fig:4}
\end{figure*}

The previous steps determined the input reduction, quadrant mapping, segments, fixed-point scale, coefficients, and squaring rule. Figure~\ref{fig:4} combines these results into an equivalent arithmetic model from the PTX input to the FP32 output.

In Fig.~\ref{fig:4}, the \texttt{FMUL.\allowbreak{}RZ} prescale belongs to the surrounding PTX operations. The bare core folds the in-period phase into the first quadrant and selects coefficients with the high 6 bits. With the low 17 bits, it evaluates the Q37 fixed-point quadratic $y=c_s+b_s\ell+a_sS_{15}(v)$, where $v=\ell\gg2$ and the quadratic term uses the truncated squaring rule above. In the linear window, $S_{15}(v)=0$, and the expression reduces to the linear model of Section~\ref{sec:3.2.4}. The result is normalized at its most significant bit and truncated to FP32. The output sign is then set from the quadrant and the input sign. Special inputs and other cases that skip the normal evaluation are handled by bypass paths. This flow is a functionally equivalent model that agrees with the measured outputs. It does not imply the physical implementation inside the chip.

\subsection{Other Functions and Data Formats}\label{sec:3.3}

Section~\ref{sec:3.2} used SIN/COS to show the full process, from input reduction and segment detection to the recovery of finite arithmetic. Applying the same observation and constraint method to the minimal PTX entries in Table~\ref{tab:1} gives the numerical specifications of the other target MUFU instructions. The results show that, in SFU-Sim, the 14 MUFU forms studied here fall into 7 numerical core families. Forms in the same family share the same segment coefficients and evaluation rules.

Every family follows the same three stages. First, the input is split into a sign, an exponent, and a mantissa. Next, an integer derived from the input goes to a piecewise quadratic core. Finally, the core output is combined with the exponent and sign to form the result. Table~\ref{tab:4} lists these stages for each family. It does not repeat the step-by-step recovery of each form.
\begin{table*}[!t]
\caption{Numerical cores of the target MUFU instructions: input split, core, result, and bypasses}\label{tab:4}
\centering\footnotesize
\setlength{\tabcolsep}{3pt}\renewcommand{\arraystretch}{1.05}
\begin{tabular}{@{}>{\raggedright\arraybackslash}p{1.02in} >{\raggedright\arraybackslash}p{1.55in} >{\raggedright\arraybackslash}p{1.5in} >{\raggedright\arraybackslash}p{1.15in} >{\raggedright\arraybackslash}p{1.4in}@{}}
\toprule
\multirow{2}{*}{\textbf{Core family}} & \multicolumn{3}{c}{\textbf{Normal finite inputs}} & \multirow{2}{*}{\textbf{Bypasses (special inputs)}}\\
\cmidrule(lr){2-4}
 & \textbf{Split the input} & \textbf{Core: segments and input $\rightarrow$ output} & \textbf{Rebuild the result} & \\
\midrule
\textbf{SIN/COS} \newline \texttt{MUFU.\allowbreak{}SIN} \newline \texttt{MUFU.\allowbreak{}COS} & The fraction of $\lvert t\rvert$ gives a 25-bit code $2^{23}q+p$: quadrant $q$ and in-quadrant phase $p$. COS first advances $q$ by one. & 6+17 bits, 64 segments \newline $u=p$ (or $2^{23}-1-p$ in odd quadrants) \newline $u\to K$ & $\pm K$; sign from $q$ and the input sign & NaN/$\pm\infty$ $\rightarrow$ NaN \newline $\pm0$/subnormal: SIN $\rightarrow$ $\pm0$, COS $\rightarrow$ $1$ \\
\midrule[\lightrulewidth]
\textbf{EX2} \newline \texttt{MUFU.\allowbreak{}EX2} \newline \texttt{MUFU.\allowbreak{}EX2.\allowbreak{}F16} \newline \texttt{MUFU.\allowbreak{}EX2.\allowbreak{}BF16}\textsuperscript{\textdagger} & $x=n+f$, $n\in\mathbb Z$, $f\in[0,1)$, $f=u/2^{23}$ & 6+17 bits, 64 segments \newline $u\to 2^f$ & $2^x=2^n\cdot2^f$ & NaN $\rightarrow$ NaN \newline $x\ge128$ $\rightarrow$ $+\infty$; $x\le-128$ $\rightarrow$ $+0$ \newline zero/subnormal $\rightarrow$ $1$ \\
\midrule[\lightrulewidth]
\textbf{TANH} \newline \texttt{MUFU.\allowbreak{}TANH} \newline \texttt{MUFU.\allowbreak{}TANH.\allowbreak{}F16} \newline \texttt{MUFU.\allowbreak{}TANH.\allowbreak{}BF16}\textsuperscript{\textdagger} & $2^{-8}<\lvert x\rvert<8$. The exponent selects one of 11 intervals. $w$ = top 17--21 mantissa bits (the width depends on the interval). & Segment index within the interval + 17 bits; 51 coefficient sets in total \newline $w\to\delta$ & $\lvert x\rvert$ with the bits below $w$ cleared, minus $\delta$; sign restored & $\lvert x\rvert\le2^{-8}$ $\rightarrow$ $x$ \newline $\lvert x\rvert\ge8$ $\rightarrow$ $\pm1$ \newline NaN $\rightarrow$ NaN \\
\midrule[\lightrulewidth]
\textbf{RCP} \newline \texttt{MUFU.\allowbreak{}RCP} \newline \texttt{MUFU.\allowbreak{}RCP64H} & $x=(-1)^{\sigma}2^e m$, $m=1+u/2^{23}$ & 7+16 bits, 128 segments \newline $u\to m^{-1}$ & $x^{-1}=(-1)^{\sigma}2^{-e}m^{-1}$ & NaN $\rightarrow$ NaN \newline $\pm\infty$ $\rightarrow$ $\pm0$ \newline zero/subnormal $\rightarrow$ $\pm\infty$ \\
\midrule[\lightrulewidth]
\textbf{RSQ} \newline \texttt{MUFU.\allowbreak{}RSQ} \newline \texttt{MUFU.\allowbreak{}RSQ64H} & $x=2^{2k+r}m$, $r\in\{0,1\}$, $m=1+u/2^{23}$ & 6+17 bits; two tables (by $r$), 64 segments each \newline $(u,r)\to(2^rm)^{-1/2}$ & $x^{-1/2}=2^{-k}(2^rm)^{-1/2}$ & NaN/$-\infty$/negative normal $\rightarrow$ NaN \newline $+\infty$ $\rightarrow$ $+0$ \newline zero/subnormal $\rightarrow$ $\pm\infty$ \\
\midrule[\lightrulewidth]
\textbf{LG2} \newline \texttt{MUFU.\allowbreak{}LG2} & $x=2^e m$, $m=1+u/2^{23}$ & 6+17 bits, 64 segments \newline $u\to\log_2 m$ & $\log_2x=e+\log_2m$ & NaN/$-\infty$/negative normal $\rightarrow$ NaN \newline zero/subnormal $\rightarrow$ $-\infty$ \newline $1$ $\rightarrow$ $+0$; $+\infty$ $\rightarrow$ $+\infty$ \\
\midrule[\lightrulewidth]
\textbf{SQRT} \newline \texttt{MUFU.\allowbreak{}SQRT} & $x=2^{2k+r}m$, $r\in\{0,1\}$, $m=1+u/2^{23}$ & 6+17 bits; two tables (by $r$), 64 segments each \newline $(u,r)\to(2^rm)^{1/2}$ & $\sqrt{x}=2^{k}(2^rm)^{1/2}$ & NaN/$-\infty$/negative normal $\rightarrow$ NaN \newline $+\infty$ $\rightarrow$ $+\infty$ \newline zero/subnormal $\rightarrow$ $\pm0$ \\
\bottomrule
\end{tabular}
\par\smallskip{\footnotesize\raggedright\textbf{Note:} \textsuperscript{\textdagger} The BF16 forms apply only to the tested Blackwell (\texttt{sm\_\allowbreak{}120}). For SIN/COS, $t$ is the bare MUFU input in turns (Section~\ref{sec:3.2.1}), and $K$ is the magnitude kernel (Section~\ref{sec:3.2.2}). $u$ is the 23-bit integer passed to the core, and $\sigma$ is the sign bit. Every FP32 NaN output is the fixed encoding \texttt{0x7fffffff}, and the input payload is not kept.\par}
\end{table*}

Each core takes the coefficients of its segment from a table. It then evaluates a piecewise quadratic with a truncated square term. FP32 results are truncated, except those of RCP and RSQ, which round to nearest even. Subnormal results of EX2 and RCP are flushed to zero.

The narrow formats reuse the FP32 cores. An FP16 input is widened to FP32, and the result is truncated toward zero. A BF16 input is placed in the high 16 bits of an FP32 word, and the high 16 bits of the result are kept. A narrow-format NaN returns \texttt{0x7fff}. RCP64H and RSQ64H pad the high 20 mantissa bits with three zeros and run the FP32 model on the result. They keep the top 20 mantissa bits of its output, adjust the exponent, and rebuild the output high word. The surrounding PTX operations set the low output word to zero.

These models only need to store segment coefficients. The coefficients of all 8 FP32 functions take less than 10 KB in total, and the FP16, BF16, and FP64 high-word forms reuse them. In contrast, storing the outputs of these 8 functions for all $2^{32}$ inputs would take about 128 GiB.

\section{Bit-Exact Validation and PTX Path Extension}\label{sec:4}

\subsection{Experimental Setup and Validation Method}\label{sec:4.1}

Section~\ref{sec:3} gave the equivalent MUFU models. This section checks the outputs of the target PTX paths on the NVIDIA A100, the GeForce RTX 4060 Laptop GPU, and the RTX PRO 5000 Blackwell. The compile targets are \texttt{sm\_\allowbreak{}80}, \texttt{sm\_\allowbreak{}89}, and \texttt{sm\_\allowbreak{}120}, respectively. The common minimal paths on all three devices use the same SFU-Sim coefficients and arithmetic rules, with no per-device refitting. The tested Blackwell also enables the two BF16 forms. For each PTX path, the split between the MUFU and the surrounding operations follows the SASS of the corresponding compiled binary. Validation first covers the minimal paths and then extends to the other tested forms.

\begin{figure*}[!t]
\centering
\includegraphics[width=0.74\textwidth]{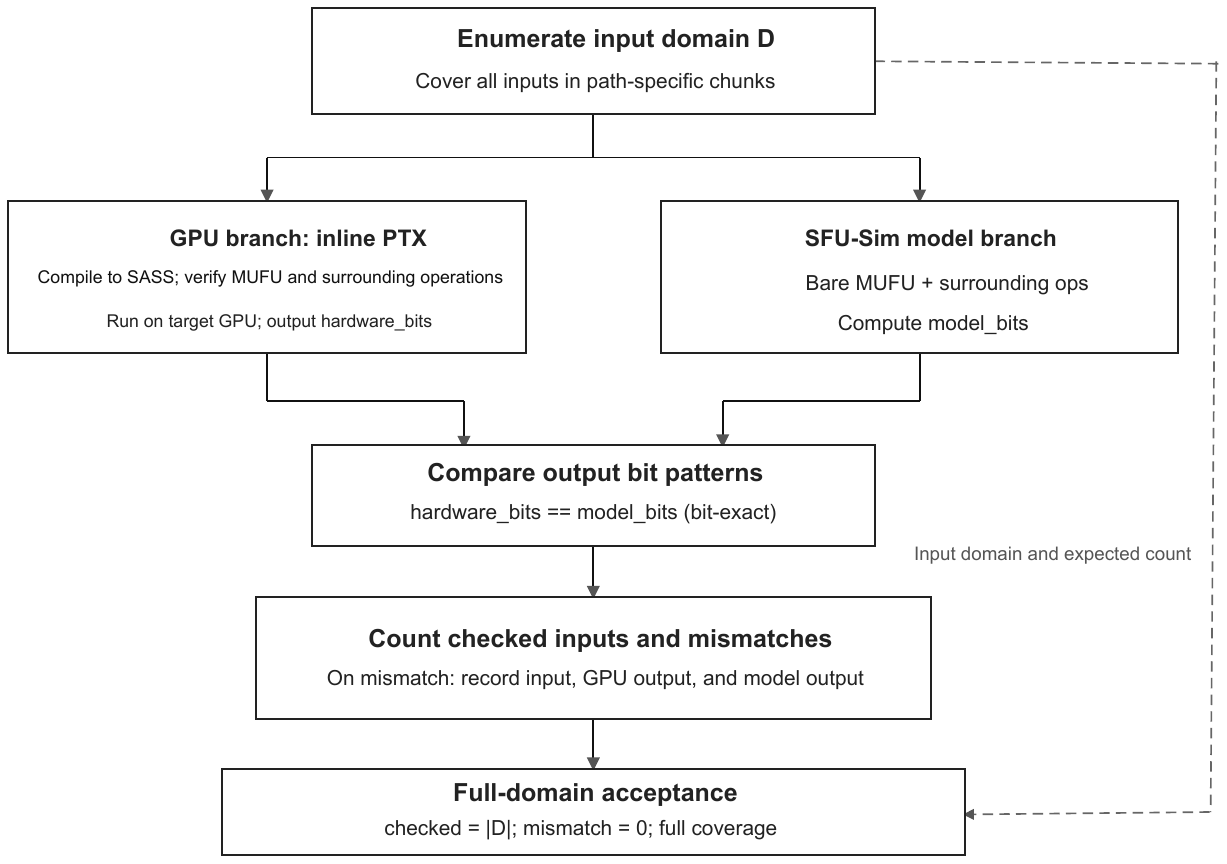}
\caption{Bit-exact validation flow. The same input is fed to the GPU and to SFU-Sim, and their output bit patterns are compared.}\label{fig:5}
\end{figure*}

The test programs specify the operation under test with inline PTX. We inspect the compiled SASS to confirm which MUFU instruction actually runs and which operations surround it. We then feed the same input to the GPU and to the independently computed SFU-Sim model, and compare the two outputs bit for bit. Figure~\ref{fig:5} shows the validation flow.

The experiments enumerate the input domain of each path as defined in Section~\ref{sec:2.3}. A path passes only when the enumeration has no gaps or duplicates, the number of checked inputs equals the domain size, and the GPU and SFU-Sim outputs have zero mismatches. We also inject a single-bit fault into the common comparator to confirm that mismatches are detected.

\subsection{Full-Domain Validation of Minimal PTX Paths}\label{sec:4.2}

We first test whether the bare models of Section~\ref{sec:3}, combined with the necessary surrounding PTX operations, reproduce the real GPU outputs. The A100, RTX 4060, and RTX PRO 5000 Blackwell are each validated on the 12 common paths. The Blackwell is also validated on the two BF16 paths. Table~\ref{tab:5} summarizes the results by format.
\begin{table*}[!t]
\caption{Bit-exact validation results of the minimal PTX paths over their specified input domains}\label{tab:5}
\centering\footnotesize
\setlength{\tabcolsep}{3pt}\renewcommand{\arraystretch}{1.1}
\begin{tabularx}{\textwidth}{@{}l l Y c c c@{}}
\toprule
\textbf{Path group} & \textbf{Functions} & \textbf{Input domain of each path} & \textbf{A100} & \textbf{RTX 4060} & \textbf{RTX PRO 5000} \\
\midrule
FP32 & EX2, LG2, RCP, RSQ, SQRT, SIN, COS, TANH (8) & All $2^{32}$ FP32 encodings & 0 & 0 & 0 \\
FP16 & EX2, TANH (2) & All $2^{16}$ FP16 encodings & 0 & 0 & 0 \\
FP64 & RCP, RSQ (2) & All $2^{32}$ high words, low word fixed to zero; full 64-bit output compared & 0 & 0 & 0 \\
BF16 & EX2, TANH (2) & All $2^{16}$ BF16 encodings & --- & --- & 0 \\
\bottomrule
\end{tabularx}
\end{table*}

Each ``0'' means that every listed path in the group has zero mismatches on that device. ``---'' means that we have no result for that case. Table~\ref{tab:1} gives the full PTX entry of each path.

Table~\ref{tab:5} shows that the 12 common minimal paths have zero mismatches over their specified input domains on all three devices. The two BF16 paths also have zero mismatches on the tested Blackwell. SFU-Sim thus reproduces the GPU outputs of these minimal PTX paths bit for bit.

\subsection{Modeling and Full-Domain Validation of Extended PTX Paths}\label{sec:4.3}

Beyond the minimal paths in Section~\ref{sec:4.2}, we further validate three groups of paths on the A100: seven non-\texttt{.ftz} FP32 approximate paths, two FP16x2 paths, and eight FP32 RCP/SQRT paths with the RN, RZ, RM, and RP rounding modes. Table~\ref{tab:6} groups them by extension type. Each path covers all $2^{32}$ input encodings, and all paths have zero mismatches.
\begin{table*}[!t]
\caption{Bit-exact validation results of the extended PTX paths on the A100}\label{tab:6}
\centering\footnotesize
\setlength{\tabcolsep}{3pt}\renewcommand{\arraystretch}{1.1}
\begin{tabular}{@{}>{\raggedright\arraybackslash}p{1.45in} >{\raggedright\arraybackslash}p{1.35in} >{\raggedright\arraybackslash}p{1.85in} >{\raggedright\arraybackslash}p{1.3in} >{\centering\arraybackslash}p{0.75in}@{}}
\toprule
\textbf{Path type} & \textbf{Functions and number of paths} & \textbf{How SFU-Sim predicts the output} & \textbf{Input domain of each path} & \textbf{Mismatches per path} \\
\midrule
Non-\texttt{.ftz} FP32 approximate & EX2, LG2, RCP, RSQ, SQRT, SIN, COS (7) & Bare model combined with the matching surrounding PTX operations & All $2^{32}$ FP32 encodings & 0 \\
FP16x2 & EX2, TANH (2) & Each FP16 half-word evaluated separately, then packed & All $2^{32}$ half-word combinations & 0 \\
FP32 RCP/SQRT with four rounding modes & Four each for RCP and SQRT (8) & Final output predicted directly from the PTX rounding semantics & All $2^{32}$ FP32 encodings & 0 \\
\bottomrule
\end{tabular}
\end{table*}

The \texttt{.ftz} forms may flush subnormal results to zero. The non-\texttt{.ftz} forms must handle ranges where the output can be subnormal. Take EX2 as an example. In \texttt{ex2.\allowbreak{}approx.\allowbreak{}ftz.\allowbreak{}f32}, the computation directly runs \texttt{MUFU.\allowbreak{}EX2}. In \texttt{ex2.\allowbreak{}approx.\allowbreak{}f32}, for finite inputs $x<-126$, an FP32 multiply first computes $x/2$ and sends it to \texttt{MUFU.\allowbreak{}EX2}. A second FP32 multiply then squares the result. Other finite inputs go directly to \texttt{MUFU.\allowbreak{}EX2}. Since $2^{-126}$ is the smallest normal FP32 number, this method uses $2^x=(2^{x/2})^2$. It first computes a larger intermediate result and then forms an output that may be subnormal. Both forms use the same \texttt{MUFU.\allowbreak{}EX2} core. The second path only adds conditional scaling and output correction. SFU-Sim must therefore reproduce both the bare \texttt{MUFU.\allowbreak{}EX2} behavior and these surrounding operations bit for bit. The other non-\texttt{.ftz} paths are handled in the same way, each with its own verified surrounding operations.

The FP16x2 paths treat the low and high half-words of the input as two independent FP16 encodings. The tested SASS uses two FP16 MUFU instructions of the same name, one for each half-word, and then packs the two 16-bit results into the output. The input domain in Table~\ref{tab:6} covers all combinations of the two half-words. It is not a scan with one half-word fixed.

The RCP/SQRT paths with four rounding modes differ from the first two types. SFU-Sim evaluates them directly from the final rounding semantics of PTX. It uses the integer quotient and remainder, or the integer root and remainder, to determine the output. It neither calls the bare MUFU model nor replays the SASS instruction by instruction. The tested hardware sequences do involve \texttt{MUFU.\allowbreak{}RCP} and \texttt{MUFU.\allowbreak{}RSQ}. Even so, the zero mismatches on these eight paths only validate the prediction of the final PTX output. They are not an independent validation of the MUFU intermediate results.

\section{Conclusion}\label{sec:5}

Special function approximation instructions have public error bounds, but these bounds cannot determine the output bit patterns of a target GPU. To address this problem, we propose SFU-Sim. By analyzing the SASS path of each PTX instruction, we separate the behavior of a single MUFU from its surrounding operations. By combining real GPU outputs with mathematical analysis, we recover the segment coefficients and the finite-width arithmetic rules. With SIN/COS as a complete case study, we show how to recover the phase reduction, the segment structure, the coefficients, and the truncated squaring rule. We also organize the 14 MUFU forms studied here into seven numerical core families.

SFU-Sim implements these rules as computable bare MUFU models and combines them with the matching surrounding operations to predict the final outputs of PTX paths. On the NVIDIA A100, the GeForce RTX 4060 Laptop GPU, and the RTX PRO 5000 Blackwell, the model outputs match the hardware bit for bit on every tested path over its specified input domain, with zero mismatches. Unlike a lookup table that stores the output for every input, SFU-Sim computes results from coefficients and arithmetic rules. It predicts the outputs of the target paths without a physical GPU, and it provides a numerical behavior model that can be analyzed. The SFU-Sim source code and validation programs are open source and can be used to reproduce our results.


\begin{thebibliography}{00}
\bibitem{r1} NVIDIA. \textit{Parallel Thread Execution ISA}, Version 8.7. CUDA Toolkit 12.8.2 Documentation, 2025. \url{https://docs.nvidia.com/cuda/archive/12.8.2/parallel-thread-execution/index.html}.
\bibitem{r2} NVIDIA. \textit{CUDA Binary Utilities}, CUDA Toolkit 12.8.2 Documentation. Instruction set reference and disassembly utilities. \url{https://docs.nvidia.com/cuda/archive/12.8.2/cuda-binary-utilities/index.html}.
\bibitem{r3} S. F. Oberman and M. Y. Siu. A High-Performance Area-Efficient Multifunction Interpolator. In \textit{Proceedings of the 17th IEEE Symposium on Computer Arithmetic}, 2005, pp. 272--279. \url{https://doi.org/10.1109/ARITH.2005.7}.
\bibitem{r4} J.-A. Piñeiro, S. F. Oberman, J.-M. Muller, and J. D. Bruguera. High-Speed Function Approximation Using a Minimax Quadratic Interpolator. \textit{IEEE Transactions on Computers}, 54(3), 2005, pp. 304--318. \url{https://doi.org/10.1109/TC.2005.52}.
\bibitem{r5} N. Cankaya. Bit-Exact AI Inference Verification Without Performance Tradeoffs. arXiv:2606.00279v2, 2026. \url{https://arxiv.org/abs/2606.00279v2}.
\bibitem{r6} P. Xie, S. Xu, Y. Wang, F. Yang, and M. Yang. Bit-Accurate Modeling of GPU Matrix Multiply-Accumulate Units: Demystifying Numerical Discrepancy and Accuracy. arXiv:2511.10909v3, 2026. \url{https://arxiv.org/abs/2511.10909v3}.
\bibitem{r7} envytools contributors. \textit{envytools}: \texttt{nvhw/\allowbreak{}sfu.\allowbreak{}c}, \texttt{nvhw/\allowbreak{}sfu\_\allowbreak{}tab.\allowbreak{}c}, and \texttt{hwtest/\allowbreak{}g80\_\allowbreak{}sfu.\allowbreak{}cc}. Source repository. \url{https://github.com/envytools/envytools}.
\end{thebibliography}
\end{document}